\documentstyle[12pt]{article}

\large
\newcommand{\be}{\begin{eqnarray}}
\newcommand{\ee}{\end{eqnarray}}
\newcommand\del{\partial}

\begin{document}
\setlength{\baselineskip}{21pt}
\pagestyle{empty}
\vfill
\eject
\begin{flushright}
SUNY-NTG-94/1
\end{flushright}

\vskip 2.0cm
\centerline{\bf The spectrum of the QCD Dirac operator and chiral random
matrix theory:}
\vskip 0.3 cm
\centerline{\bf the threefold way.}
\vskip 2.0 cm
\centerline{Jacobus Verbaarschot}
\vskip .2cm
\centerline{Department of Physics}
\centerline{SUNY, Stony Brook, New York 11794}
\vskip 2cm

\centerline{\bf Abstract}
We argue that the spectrum of the QCD Dirac operator near zero virtuality can
be described by random matrix theory. As in the case of classical random matrix
ensembles of Dyson we have three different cases: the chiral orthogonal
ensemble
(chGOE), the chiral unitary ensemble (chGUE) and the chiral symplectic
ensemble (chGSE). They correspond to gauge groups $SU(2)$ in the
fundamental representation, $SU(N_c), N_c \ge 3$ in the fundamental
representation, and gauge groups
for all $N_c$ in the adjoint representation, respectively. The joint
probability density reproduces Leutwyler-Smilga sum rules.

\vfill
\noindent
\begin{flushleft}
SUNY-NTG-94/1\\
January 1994
\end{flushleft}
\eject
\pagestyle{plain}

\vskip 1.5cm
\setcounter{equation}{0}
\noindent
According to the Banks-Casher formula \cite{BANKS-CASHER-1980},
the spectrum of the Dirac operator
near zero virtuality is directly connected with a nonzero value of the
chiral condensate, the order parameter for the chiral phase transition.
This suggests that the spectrum in this region plays an important role
in understanding the mechanism of chiral symmetry
breaking.
Recently, it was shown that a nonzero value of the chiral condensate
leads to the existence of sum rules for inverse powers of the eigenvalues of
the Dirac operator \cite{LEUTWYLER-SMILGA-1993}. These sum rules are only
sensitive to the spectrum near zero virtuality and can be expressed into
microscopic spectral correlation functions, which measure correlations
on the order of a finite number of average level spacings
\cite{SHURYAK-VERBAARSCHOT-1993,VERBAARSCHOT-ZAHED-1993}. As is well known
from the study of chaotic systems
\cite{BOHIGAS-GIANNONI-1984,SELIGMAN-VERBAARSCHOT-ZIRNBAUER-1984},
such correlations are independent of the details of the interactions and can
be described by random matrix theory. Indeed, we
\cite{SHURYAK-VERBAARSCHOT-1993,VERBAARSCHOT-ZAHED-1993} have shown that
for of a complex Dirac operator $all$ Leutwyler-Smilga sum rules follow
from a chiral random matrix theory.
This led to the claim that the microscopic
spectral correlation functions of the Dirac operator are universal.

More than three decades ago Dyson \cite{DYSON-1962} found
three distinct
types of random matrix ensembles: the gaussian orthogonal ensemble (GOE),
the gaussian unitary ensemble (GUE) and the gaussian symplectic ensemble (GSE),
corresponding
to real, complex and quaternion matrix elements. The random matrix theory
discussed in \cite{SHURYAK-VERBAARSCHOT-1993,VERBAARSCHOT-ZAHED-1993}
has complex matrix elements. For that reason it has been named
it the chiral unitary ensemble (chGUE).
This raises the question of what are
the chiral analogues of the GOE and the GSE, and what is the structure of the
corresponding Dirac operator. The answer to this question will be given
in this letter. We also will derive the joint eigenvalue density of the
random matrix ensembles and present the result for the simplest
Leutwyler-Smilga sum rule.

The Euclidean Dirac operator in QCD is defined by
\be
D \equiv i\gamma \del + \gamma A,
\ee
where $A$ is an $SU(N_c)$ valued gauge field ($N_c$ is the number of colors).
Because this operator
anti-commutes with $\gamma_5$, in a chiral basis it reduces to the following
block structure
\be
\left ( \begin{array}{cc}
             0 & T \\ T^\dagger & 0
\end{array} \right ).
\ee
In general (for $N_c \ge 3$), if $A$ is in the the fundamental representation,
the matrix elements of the Dirac operator, $i.e.$ $T_{ij}$, are complex.
This defines the first family of Dirac operators.

However,
in the case of $SU(2)$ we have an additional symmetry
\cite{LEUTWYLER-SMILGA-1993}, which is specific
to this group:
\be
[C^{-1} \tau_2 K, D] =0,
\ee
where $C$ is the charge conjugation operator ($\gamma_\mu^* = - C\gamma_\mu
C^{-1})$, and $K$ denotes the complex conjugation operator.
This symmetry operator has the property that
\be
(C^{-1} \tau_2 K)^2 = 1.
\ee
As is well known from the analysis of the time-reversal operator
in random matrix theory \cite{PORTER-1965},
this property allows us to choose a basis in which the matrix elements
of the Dirac operator in (2) are real and $T^\dagger =
\tilde T$. This provides us with the second family of Dirac operators.

The third family of gauge theories are those with the fermions in the
adjoint representation. The Dirac operator is given by
\be
 D_{kl} = i\gamma \del \delta_{kl} + f^{klm} \gamma A_m,
\ee
where $f^{klm}$ are the structure constants of the gauge group. As was noted
in \cite{LEUTWYLER-SMILGA-1993}, in this case the covariant derivative
is real, and the Dirac operator is invariant under charge conjugation
\be
[D, C^{-1}K] =0.
\ee
Because $C^* C^{-1} = -1$, one can easily derive that
\be
\left ( C^{-1} K \right )^2 = -1.
\ee
This implies that each eigenvalue of the Dirac operator is doubly degenerate
with linearly independent eigenfunctions \cite{PORTER-1965} given by
\be
\phi_\lambda \quad {\rm and} \quad C^{-1} K \phi_\lambda.
\ee
As follows from a discussion  by Dyson \cite{DYSON-1961}, in this case
the Dirac operator can be diagonalized by a symplectic transformation. Or,
in other words, the matrix elements of $T$ can be regrouped into
real quaternions, and $T^\dagger_{ij} = \bar T_{ji}$ (quaternion conjugation
is denoted by a bar).

The QCD partition function for $N_f$ flavors with masses $m_f$
($m_f \rightarrow 0$) in the sector with $\nu$ zero modes is defined by
\be
Z_\nu^{\rm QCD} = \langle \prod^{N_f}_{f=1}\prod_{\lambda_n>0}
(\lambda_n^2 + m_f^2) m_f^{\nu}\rangle_{S_\nu(A)},
\ee
where the average $<\cdots>_{S_\nu(A)}$ is over gauge field configurations
with $\nu$ fermionic zero modes weighted by the gauge field action
$S_\nu(A)$. The product is over all eigenvalues of the
Dirac operator. For fermions in the adjoint representation, the doubly
degenerate eigenvalues count only once (Majorana fermions)
\cite{LEUTWYLER-SMILGA-1993}.
Relevant observables are obtained by differentiation
with respect to the masses. The distribution of the eigenvalues of
the Dirac operator is induced by the fluctuations of the gauge field.
It is our claim that the correlations between eigenvalues near zero
virtuality, $i.e.$ only a finite number of level spacings away from zero,
do not depend on details of the interaction.

The basic underlying idea of random matrix theory is that correlations between
eigenvalues on the scale of one eigenvalue are only determined by the
symmetries
of the system
and do not depend on the detailed dynamics. Therefore, the spectral density
measured in units of the average spectral density near zero virtuality is
a universal quantity that can be described by a random matrix theory
that reflects only on the $symmetries$ of the Dirac operator.
The relevant random matrix theory in the sector with $\nu$ zero modes is
\be
Z_{\beta,\nu} = \int {\cal D}T P_\beta(T)\prod_f^{N_f}\det \left (
\begin{array}{cc} m_f & iT\\
                 iT^\dagger & m_f
\end{array} \right ),
\ee
where $T$ has the symmetries of the corresponding Dirac operator and the
masses are in the chiral limit ($m_f\rightarrow 0$).
For $SU(2)$ in the fundamental representation $T$ is real ($\beta = 1$),
for $SU(N_c)$,
$N_c \ge 3$, the matrix $T$ is complex ($\beta = 2$),
and for fermions in the adjoint
representation, $T$ is quaternion real ($\beta = 4$). In the latter case
the square root of the fermion determinant appears in (10).
The matrix $T$ is a rectangular $n\times m$ matrix with $|n-m| = \nu$.
It can be shown  that the matrix in the fermion determinant in (10)
has exactly $\nu$ zero eigenvalues (and $Z \sim \prod_f m_f^\nu$).
The distribution function of
the matrix elements $P(T)$ that is consistent with no additional information
input is gaussian \cite{BALIAN-1968}. In a standard normalization we choose
\be
P_\beta(T) = \exp\left (
{-\frac{\beta n}{2\sigma^2}\sum_{k=1}^n \lambda_k^2}\right),
\ee
where the sum is over the non-zero eigenvalues of $T$.
With this choice the average spectral density does not depend on $\beta$ and
is given by $\rho(0) = 1/\pi\sigma$. This allows us to identify $\sigma =
1/\pi\rho(0)\equiv 1/\Sigma$, where $\Sigma$ denotes the chiral condensate.
It should be noted that as in the QCD partition function the chiral symmetry
is broken isotropically in flavor space.

In order
to derive the joint eigenvalue density we use the eigenvalues and eigenangles
of $T$ as new integration variables. In each of the three cases, up to a
constant,  the Jacobian of this transformation is given by
\be
J = \prod_{k,l} |\lambda_k^2 -\lambda_l^2|^\beta \prod_{k}
\lambda_k^{\nu\beta+\beta -1}.
\ee
The derivation of this result will be given elsewhere.
At the moment we only remark that the total powers of $\lambda_k$
can be obtained on dimensional grounds only.
The joint eigenvalue density is therefore given by
\be
\rho_\beta(\lambda_1, \cdots, \lambda_n) = C_{\beta, n}
\prod_{k,l} |\lambda_k^2 -\lambda_l^2|^\beta \prod_{k}
\lambda_k^{2N_f +\beta\nu+\beta -1}
e^{-\frac{n\beta\Sigma^2}{2} \sum_k \lambda^2_k}.
\ee
The normalization constant is denoted by $C_{n,\beta}$.
In the case of $\beta =4$, each of the doubly degenerate eigenvalues is
counted only once. This is consistent with the fact that fermions in
the adjoint representation are Majorana fermions (see
\cite{LEUTWYLER-SMILGA-1993} for a detailed discussion of this point).
The simplest Leutwyler-Smilga sum rule in the sector
with $\nu$ zero modes can be evaluated with the help of Selberg's integral
\cite{SELBERG-1944}
(see \cite{MEHTA-1991} for a discussion). The result is
\be
\left \langle\frac 1{N^2}\sum_{\lambda_k>0}
\frac 1{\lambda^2_k}\right\rangle_{\rho_\beta} = \frac
{\beta\Sigma^2}{8(\frac{\beta\nu}2+\frac \beta{2}+ N_f-1)},
\ee
where the average is with respect to the spectral density
$\rho_\beta(\lambda_1,\cdots, \lambda_n)$. The total number of
modes is denoted by $N = m+n$.
This constitutes the final result of this letter. It agrees with sum rules
obtained by Leutwyler and Smilga for $\beta = 2$ and $\beta = 4$.
This sum-rule can also be expressed in terms of the microscopic spectral
density defined by
\be
\rho_S(x) = \lim_{N\rightarrow \infty} \frac 1N \rho(\frac xN),
\ee
where $\rho(\lambda)$ is obtained from (13) by integrating over all eigenvalues
except one.
The microscopic spectral density has been derived for $\beta =1$
\cite{VERBAARSCHOT-1994} and $\beta =2$ \cite{VERBAARSCHOT-ZAHED-1993},
and in both cases it agrees with numerical results from simulations of gauge
field configurations by a liquid of instantons \cite{VERBAARSCHOT-1994}.

In \cite{LEUTWYLER-SMILGA-1993} sum rules were derived from the static limit
an effective field theory. The above discussed triality
implies that we have three
structurally different effective field theories. Two of them were analyzed
in \cite{LEUTWYLER-SMILGA-1993}, and the theory for $\beta =1$ which involves
both baryons and mesons, was discussed in
\cite{DIAKONOV-PETROV-1990}. It would be instructive to derive (14) for $\beta=
1$ in the framework of this model.

In conclusion, we have argued that, depending on the gauge group,
the QCD Dirac operator falls into three
different families: $SU(2)$ in the
fundamental representation, $SU(N_c)$, $N_c \ge 3$ in the fundamental
representation and $SU(N_c)$ in the adjoint representation.
This triality corresponds to real, complex and quaternion matrix elements.
Its spectrum near zero virtuality reflects only the symmetries of the system
and can be described in terms of chiral random matrix theory:
the chGOE, the chGUE and the chGSE, respectively. Sum rules obtained
general arguments based on effective field theory
\cite{LEUTWYLER-SMILGA-1993} are reproduced.

\vglue 0.6cm
{\bf \noindent  Acknowledgements \hfil}
\vglue 0.4cm
The reported work was partially supported by the US DOE grant
DE-FG-88ER40388.
I would like to thank M.A. Nowak for useful discussions and a critical
reading of the manuscript. T. Schaefer is thanked for stimulating discussions
and E. Shuryak for a critical reading of the manuscript.

\vfill
\eject
\newpage
\setlength{\baselineskip}{15pt}

\bibliographystyle{aip}

\begin{thebibliography}{10}

\bibitem{BANKS-CASHER-1980}
T.~Banks and A.~Casher,
\newblock Nucl. Phys. {\bf B169} (1980) 103.


\bibitem{LEUTWYLER-SMILGA-1993}
H.~Leutwyler and A.~Smilga,
\newblock Phys. Rev. {\bf D46} (1992) 5607.

\bibitem{SHURYAK-VERBAARSCHOT-1993}
E.~Shuryak and J.~Verbaarschot,
\newblock {\it Random matrix theory and spectral sum rules for the Dirac
 operator in QCD}, Nucl. Phys. {\bf A} (1993) (in press).

\bibitem{VERBAARSCHOT-ZAHED-1993}
J.~Verbaarschot and I.~Zahed, Phys. Rev. Lett. {\bf 70} (1993) 3852.

\bibitem{BOHIGAS-GIANNONI-1984}
O.~Bohigas, M.~Giannoni, \newblock
in '{\it Mathematical and computational methods in
nuclear physics}', J.S. Dehesa et al. (eds.), Lecture notes in Physics {\bf
209}, Springer Verlag 1984, p. 1;
O.~Bohigas, M.~Giannoni, and C.~Schmit,
\newblock Phys.Rev.Lett. {\bf 52}, 1 (1984).

\bibitem{SELIGMAN-VERBAARSCHOT-ZIRNBAUER-1984}
T.~Seligman and J.~Verbaarschot and M.~Zirnbauer,
\newblock Phys. Rev. Lett. {\bf 53}, 215 (1984).

\bibitem{DYSON-1962}
F.J. Dyson,
\newblock J. Math. Phys. {\bf 6} (1962) 1199.

\bibitem{PORTER-1965}
C.E.~Porter, \newblock
'{\it Statistical theories of spectra: fluctuations}', Academic
Press, 1965.

\bibitem{DYSON-1961}
F.J. Dyson,
\newblock J. Math. Phys. {\bf 3} (1962) 140.

\bibitem{BALIAN-1968}
R.~Balian,
\newblock Nuov. Cim. {\bf 57} (1968) 183.

\bibitem{SELBERG-1944}
A. Selberg,
\newblock Norsk Mat. Tid. {\bf 26} (1944) 71.

\bibitem{MEHTA-1991}
M.~Mehta,
\newblock {\it Random Matrices}, Academic Press, San Diego, 1991.

\bibitem{VERBAARSCHOT-1994}
J. Verbaarschot,
\newblock{\it The spectrum of the Dirac operator near zero virtuality for
$N_c=2$}, Stony Brook preprint NTG-94/2

\bibitem{DIAKONOV-PETROV-1990}
D. Diakonov and V. Petrov,
\newblock Bad Honnef Conference, (ed. K. Goeke), Springer, 1992.

\bibitem{VERBAARSCHOT-1994}
J. Verbaarschot,
\newblock{\it Chiral random matrix theory and the spectrum of the Dirac
operator near zero virtuality}, Acta Pol. Phys. (1994) (in press);
J.Verbaarschot, (to be published).

\end{thebibliography}

\end{document}